\documentclass[a4paper,twocolumn,abstract,titlepage=false,DIV=16,BCOR=0pt,headinclude=true]{scrartcl}
\usepackage[headsepline]{scrlayer-scrpage}
\usepackage{algorithmic}
\usepackage{algorithm}
\usepackage{array}
\usepackage{textcomp}
\usepackage{stfloats}
\usepackage{url}
\usepackage{verbatim}
\usepackage{graphicx}
\usepackage{xcolor}
\usepackage[numbers]{natbib}
\usepackage{tabularx}
\usepackage{multirow}
\usepackage{tabularray}
\usepackage{enumitem}
\setlist{itemsep=0.1\baselineskip}

\makeatletter
\g@addto@macro{\UrlBreaks}{\UrlOrds}
\makeatother

\usepackage[T1]{fontenc}

\usepackage{hyperref}

\hypersetup{
  hidelinks,
  colorlinks=true,
  allcolors=black,
  pdfstartview=Fit,
  breaklinks=true
}

\usepackage[all]{hypcap}
\usepackage[]{subcaption}

\usepackage[capitalise,nameinlink,noabbrev]{cleveref}

\crefname{listing}{Listing}{Listings}
\Crefname{listing}{Listing}{Listings}
\crefname{lstlisting}{Listing}{Listings}
\Crefname{lstlisting}{Listing}{Listings}

\usepackage{xspace}
\usepackage[group-minimum-digits=4,per-mode=fraction]{siunitx}

\usepackage[english]{babel}

\usepackage[
  babel=true, 
  expansion=false,
  protrusion=alltext-nott, 
  final 
]{microtype}

\title{\vspace{-2.5\baselineskip}Fingerprint Theft Using Smart Padlocks:\\ Droplock Exploits and Defenses}

\author{Steve Kerrison\\James Cook University, Singapore Campus\\\href{mailto:steve.kerrison@jcu.edu.au}{steve.kerrison@jcu.edu.au}}
\date{\today}

\ihead{S. Kerrison}
\ohead{Fingerprint Theft Using Smart Padlocks: Droplock Exploits and Defenses}

\begin{document}

\nocite{KerrisonDroplocksiThings2022}

\def\droplock{{\em droplock}\xspace}
\def\Droplock{{\em Droplock}\xspace}
\def\droplocks{{\em droplocks}\xspace}
\def\Droplocks{{\em Droplocks}\xspace}

\newcommand{\PreserveBackslash}[1]{\let\temp=\\#1\let\\=\temp}
\newcolumntype{C}[1]{>{\PreserveBackslash\centering}p{#1}}
\newcolumntype{R}[1]{>{\PreserveBackslash\raggedleft}p{#1}}
\newcolumntype{L}[1]{>{\PreserveBackslash\raggedright}p{#1}}

\newenvironment{onecolabstract}{}{}
\twocolumn[%
\maketitle
\vspace{-2.0\baselineskip}
\begin{onecolabstract}
\begin{center}
  \noindent\rule{0.45\linewidth}{1pt}
  \vspace{\baselineskip}\vfill
  \begin{minipage}{0.9\linewidth}
  There is growing adoption of smart devices such as digital locks with remote control and sophisticated authentication mechanisms. However, a lack of attention to device security and user-awareness beyond the primary function of these IoT devices may be exposing users to invisible risks.
  This paper extends upon prior work that defined the ``\droplock'', an attack whereby a smart lock is turned into a wireless fingerprint harvester. We perform a more in-depth analysis of a broader range of vulnerabilities and exploits that make a \droplock attack easier to perform and harder to detect. Analysis is extended to a range of other smart lock models, and a threat model is used as the basis to recommend stronger security controls that may mitigate the risks of such as attack.
  \end{minipage}
  \vfill
  \vspace{\baselineskip}
  \noindent\rule{0.45\linewidth}{1pt}
  \vspace{\baselineskip}
\end{center}
\end{onecolabstract}
]


\section{Introduction}
\label{sec:introduction}

Biometrics are a commonly used authentication factor, particularly on smartphones, but also on smart devices such as physical locks. However, the cost and sophistication of the devices, and the degree to which they protect fingerprint data, varies significantly. A smartphone, which may contain banking applications that require biometric authentication, has far more security controls in place than a smart padlock.

In information security, prevailing advice is to avoid using the same password on multiple platforms, in case a compromise of one platform leads to credential reuse on another. We cannot do the same with biometric data, however, as we have a limited number of fingers. This leads to a potential problem: using a compromised or counterfeit device to steal fingerprints, then use them against a victim on other systems.

This issue was initially explored in~\cite{KerrisonDroplocksiThings2022}, where the ``IoT \droplock'' concept was introduced. A \droplock is a smart padlock with fingerprint reader, left to be encountered by potential victims, who may interact with its fingerprint scanner, unknowingly having their biometrics wirelessly transmitted to a nearby attacker. The prior work focuses on turning a single candidate device into a \droplock by reverse engineering it and replacing its firmware. This work seeks to extend those contributions with further exploration of the issue as a concept, a broader survey of devices, and more detailed recommendations for security controls.

\subsection{Contributions}
\label{sec:contributions}

Expanding upon~\cite{KerrisonDroplocksiThings2022}, in this paper we make the following new contributions:

\begin{enumerate}

  \item Comprehensive details of all vulnerabilities exploited in the making of a \droplock from a COTS device.
  \item A survey of five COTS smart padlocks, to understand the broader security posture of this class of devices.
  \item Investigation into and discussion of the feasibility of producing a convincing self-made \droplock.
  \item Expanding the best practice and security controls recommendations in light of further investigations, supported by new threat modelling of the scenario.

\end{enumerate}
In combination, these contributions highlight a gap in the understanding of the severity of weak security in devices that do not themselves necessarily need it, but risk compromising immutable authentication factors that may be used elsewhere in more secure systems.

\subsection{Structure}
\label{sec:structure}

We refer to the prior work detailed in~\cite[\S~IV]{KerrisonDroplocksiThings2022} as the basis for the related work of this piece. In \Cref{sec:remote}, extensive details of additional hacks of the original \droplock target device are given, with recommendations made for improving the device- and implementation-specific vulnerabilities that were found. In \cref{sec:otherdevices}, we examine a wider range of commercially available smart padlocks to gain an impression of general susceptibility of devices to \droplock attacks, and explore the feasibility of custom-building a practical \droplock. We present threat model of the \droplock scenario in \cref{sec:recommendations}, and use it to support security control recommendations. \Cref{sec:conclusion} summarizes and concludes this paper along with discussion of potential future work.

\section{Wireless takeover of COTS locks}
\label{sec:remote}

\begin{figure*}
  \centering
  \subcaptionbox{Original cloud-dependent implementation\label{fig:enrol-orig}}{\includegraphics[height=21.00cm,trim={0mm 0mm 0mm 0mm},clip]{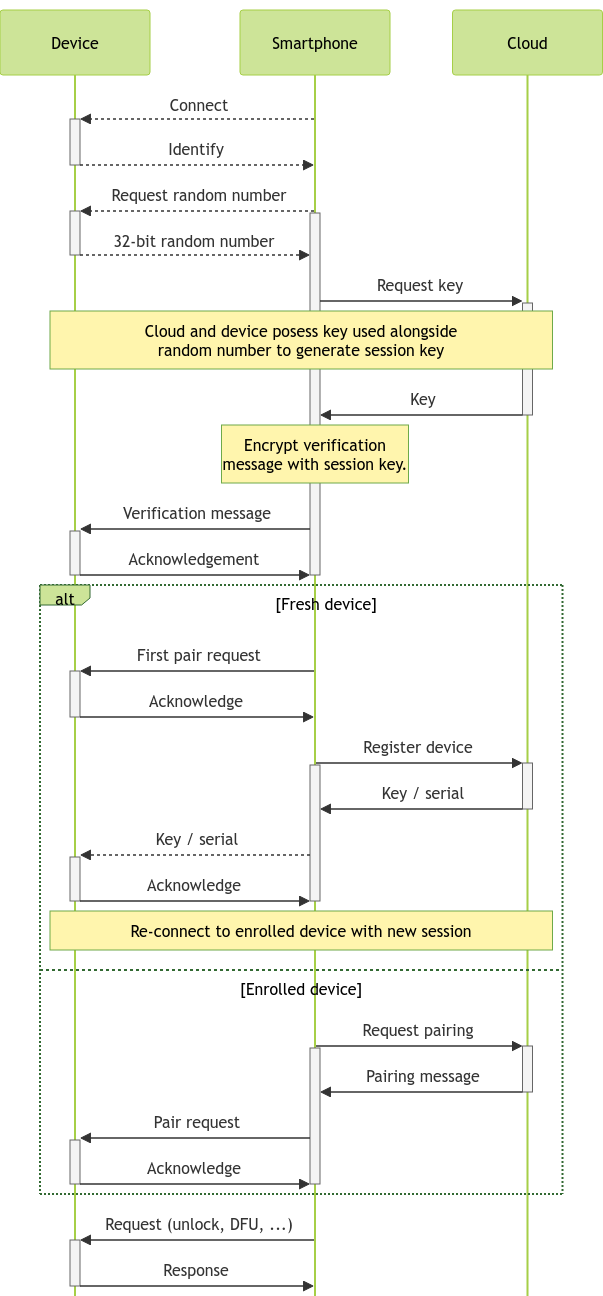}}
  \hspace{0.125cm}{\color{gray}\vrule}\hspace{0.125cm}
  \subcaptionbox{Hacked version\label{fig:enrol-hack}}{\includegraphics[height=19.78cm,trim={0mm 0mm 0mm 0mm},clip]{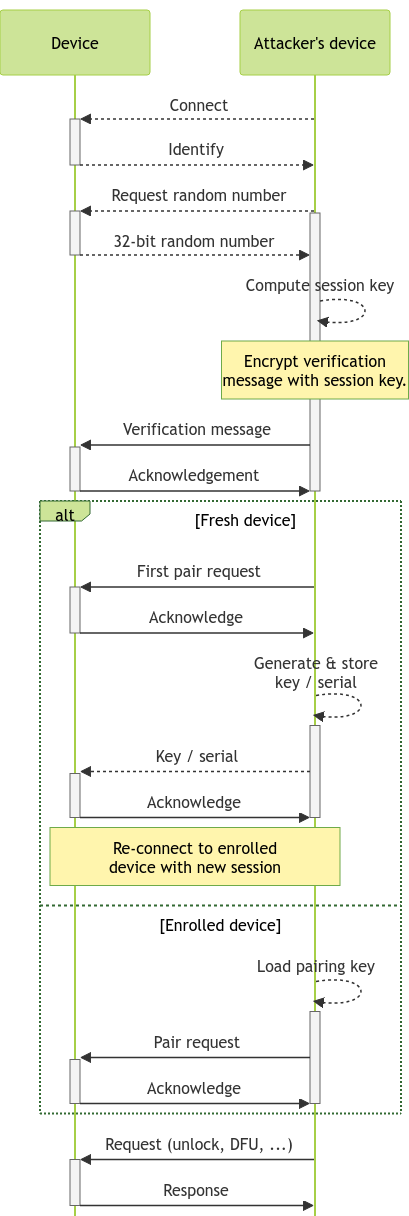}\vspace{1.211cm}}
  \caption{Sequence diagrams of exchanges required to establish session encryption with a TLL05A and enrol/control it.}
  \label{fig:enrol}
\end{figure*}

Prior work~\cite{KerrisonDroplocksiThings2022} demonstrates the proof-of-concept and how this concept can be applied to reverse-engineered commercial devices, with appropriate physical access. Remote takeover via Over The Air (OTA) Device Firmware Update (DFU) was also shown in limited detail. This was performed on a paricular lock model, the TLL05A. In this section, continuing analysis of the same lock, we expand upon a series of vulnerabilities and associated exploits that enabled brand-new locks to be reprogrammed as \droplocks without physical disassembly or registration. Other lock models are considered in \Cref{sec:otherdevices}.

\subsubsection*{Disclosure of vulnerabilities}

For the lock targeted in this section of our work, we disclosed all related vulnerabilities to the manufacturer via their bug bounty program on or before 5th May 2022. We received no non-automated response from them prior to presenting our initial findings in August 2022, and at the time of producing this extended work have yet to receive anything further. As this work focuses on more in-depth details, and the general applicability of them to this class of device, we have opted not to disclose disassembly of the manufacturer's code and instead discuss the tools and approaches that were required to successfully mount an attack, which would be repeatable by suitably skilled individuals.

\begin{table*}
  \begin{center}
  \caption{Defenses, their vulnerabilities and exploits for the target device and associated Android app.}
  \label{tab:defvulnexpl}
  \small
  \begin{tblr}{
    width = \linewidth,
    colspec = {Q[90,m]Q[30,m]Q[120,m]Q[120,m]Q[120,m]Q[120,m]},
    column{1} = {r},
    column{2} = {c},
    column{3-6} = {l},
    cell{2}{1} = {r=4}{},
    cell{2}{3} = {r=2}{},
    cell{4}{3} = {r=3}{},
    cell{4}{5} = {r=3}{},
    cell{6}{1} = {r=4}{},
    cell{7}{6} = {r=2}{},
    vlines,
    hline{1-2,10} = {-}{},
    hline{3} = {2,4-6}{},
    hline{4,7,9} = {2-6}{},
    hline{5} = {2,4,6}{},
    hline{6} = {1-2,4,6}{},
    hline{8} = {2-5}{},
  }
  \textbf{Component} & \textbf{Ref} & \textbf{Objective}                                         & \textbf{Defense}                                   & \textbf{Vulnerability}         & \textbf{Exploit / evade}               \\
  Mobile app         & \labelcref{def:pinning} & Intercept data between app and cloud API                   & \nameref{def:pinning}                                & Patchable                      & Frida patching                 \\
                     & \labelcref{def:apienc} &             & \nameref{def:apienc}                             & Static keys                    & Key extraction                 \\
                     & \labelcref{def:proguard}           & Identify device to app comms and algorithms                & \nameref{def:proguard}                                   & Reverse engineering            & JADX, manual effort            \\
                     & \labelcref{def:striplib}           &                                                            & \nameref{def:striplib}                         &                                & Ghidra, manual effort          \\
  Device             & \labelcref{def:rawfirmware}           &                                                            & \nameref{def:rawfirmware} &                                & Ghidra, debug, manual effort   \\
                     & \labelcref{def:enrollment}           & Onboard device without original app or cloud API           & \nameref{def:enrollment} & Hard-coded key                 & Reverse engineer, re-implement \\
                     & \labelcref{def:session}           & Establish device session without original app or cloud API & \nameref{def:session} & Derived from observable values &                                \\
                     & \labelcref{def:dfu}           & Update device with malicious firmware                      & \nameref{def:dfu} & No signature check             & Generate new checksum
  \end{tblr}
  \end{center}
\end{table*}

\renewcommand\thesubsubsection{\Alph{subsubsection}}
\subsection{Attacks and failed defenses}
\label{sec:attacks}

In an ideal attack, any candidate device could be wirelessly converted into a \droplock. While the target lock possesses a number of defenses against some, many can be defeated to create an almost-ideal candidate. \Cref{tab:defvulnexpl} summarizes the defenses and how they were overcome, which will be explained in further detail in this section, where each ID that is discussed matches its subsection label (A, B, etc.).

\Cref{fig:enrol} provides a sequence of communications between the target device and its smartphone and cloud API (\cref{fig:enrol-orig}), or attacker after circumventing the cloud-dependent mechanisms (\cref{fig:enrol-hack}). Dashed lines indicate unencrypted communication, while solid lines are encrypted, either by TLS and additional payload encryption for the case of smartphone to cloud communication, or session encryption of data within BLE payloads to and from the smart lock.

\subsubsection{Certificate pinning}
\label{def:pinning}

Applications that use SSL/TLS must trust one or more root Certificate Authorities (CAs) in order to verify the identity and validity of any server certificate that is presented to them. This can be provided by the Operating System (OS) or be application specific. Additionally, certificate pinning can be used to restrict allowed certificates for a host to a pre-defined list~\cite{BhorPinning}. This assumes that app updates will be frequent enough to accommodate any future certificate updates.

Certificate pinning frustrates analysis of an app's API calls by preventing the use of a Man-in-the-Middle (MitM) proxy and injected CA that would otherwise be seen as valid to the phone's OS. The app will still not validate the MitM certificate as it is not one of those predefined. This approach is discouraged by best practices documents, for example for Android developers~\cite{AndroidPinning}, due to manageability problems that can arise from its use.

Tools such as Frida~\cite[p. 37]{KotipalliSrinivasaRao2016HA} can be used to unpack and repack applications to remove or replace their pinning~\cite[pp 184--186]{KotipalliSrinivasaRao2016HA}. Although this can also be thwarted by use of customized TLS libraries, the effort involved may deter going to such lengths. In the case of the target lock, Frida was able to repack the application. This made it possible to intercept and decrypt the HTTPS traffic between the lock's phone app and the vendor's cloud API.

\subsubsection{API payload encryption}
\label{def:apienc}

Despite defeating defense \labelcref{def:pinning}, the app and API server also encrypt the majority of HTTP payloads and responses. However, the key is static and stored within the application binary using the \texttt{Cipher.so} library, which is known to be reversible~\cite{libcipher-weak}.

During packet capture, it was noted that identical payloads were sometimes received and that the payload sizes were aligned to 16 byte boundaries. This suggested that AES-ECB encryption was being used, which in itself is considered weak in the majority of use cases~\cite[p. 228]{HAC}.

The secrets protected by \texttt{Cipher.so} were extracted and one of them was found to be a key able to decrypt the captured AES-ECB payloads. This allowed observing the app and server's traffic, as well as writing an independent client able to communicate with the cloud service.

\subsubsection{Java obfuscation}
\label{def:proguard}

Unpacking the APK revealed code that appears to have been obfuscated with ProGuard~\cite{ProGuard}. This obfuscation tool employs layers of indirection and symbol renaming in order to make reverse engineering harder. However, using a tool such as JADX allows variable and method renaming, which in combination with inspection of the code and observation of interactions with system libraries whose symbols cannot be renamed, can lead to adequate reverse engineering results.

\subsubsection{Stripped C / Rust libraries}
\label{def:striplib}

Despite being able to reverse engineer parts of the app, the communication between the lock and app was implemented in a shared object containing code generated from C and Rust. Ghidra~\cite{Ghidra} was used to examine this to some extent, although at the time of writing Ghidra does not have much support for analyzing Rust code, increasing the time taken to reverse engineer some data structures. We note that if Rust continues to gain popularity in memory-safe embedded development, then analysis of Rust binaries will become more important in security testing and malware analysis.

\subsubsection{Stripped binary}
\label{def:rawfirmware}

As is expected on a low-memory embedded device, the firmware image extracted from the target lock (or obtained from the decrypted cloud API), is stripped of all debug symbols and has very few meaningful symbol names in the symbol table.

Reverse engineering of this binary required by-hand renaming of functions and variables, in conjunction with real-time debugging of the device. Stepping through the firmware with a debug device sped up the identification of the most interesting code segments, and while still time-consuming, did eventually reveal the enrollment process, session key derivation algorithm and presence of standard DFU capabilities.

\subsubsection{Enrollment key}
\label{def:enrollment}

When a target lock is purchased, the owner must register it with the vendor's cloud via phone app. During this process, a hard-coded key is used to establish a session between the lock and app, then a new unique pair of serial and key are assigned to the lock. These are used for the derivation of all future sessions and ordinarily are known only to the lock and the vendor's servers.

In combination with the reverse engineering of the enrollment process, a script was developed that could use the hard-coded key to enrol a lock with a key and serial chosen by the attacker, with no cloud API interaction necessary. The script is available via this work's accompanying data publication~\cite{KerrisonDroplockData}. This means a purchased lock is not known by the vendor as activated, which improves secrecy for the attacker, and means the attacker doesn't need to have a registered user account.

Although this attack only works on fresh, unenrolled locks, it may be possible to perform this attack during legitimate enrollment. If the attacker is in range of a target lock while it is being enrolled, the lower latency of a locally-implemented enrollment process may beat the legitimate cloud-based process. The device would appear unusable to the original owner, however, so this attack is not particularly useful.

\subsubsection{Ephemeral session keys}
\label{def:session}

Normal communication with between the lock and app \emph{requires} cloud connectivity in order to establish a session key. The lock and app use a standard BLE UART interface, without any encryption at the Bluetooth layer. Instead, AES-ECB is used after establishing the session key. This is performed by requesting a random number from the lock, which is then provided to the cloud, itself returning the session key that the app can then use to perform AES-ECB.

While the cloud is used to keep knowledge of the key derivation from the app and any observer of the app's traffic, the lock must also implement the derivation function. The device reverse engineering allowed this to be re-implemented on a local PC. It requires the serial, a separate key and the random number. The first two can be obtained from the cloud API and the third is provided on request by the lock over BLE.

A registered user can use the exploit against defense \labelcref{def:apienc} to obtain the serial and key, then run a local script to interact directly with the lock thereafter. The exploit against defense \labelcref{def:enrollment} allows an unregistered attacker to do the same.

\subsubsection{DFU checksum}
\label{def:dfu}

Nordic's standard DFU mechanism is used on the target lock. However, activation of DFU mode is performed within the encrypted session between lock and app. A combination of BLE packet capture, payload decryption and library reverse engineering revealed the command sequence needed to activate it.

Beyond this, the protections are minimal, as PKI signing is not used to protect the integrity of firmware updates. The firmware package metadata obtained via the cloud API showed that the firmwares were only CRC16 checksummed, with no signatures or chain of trust. Once a viable replacement firmware was developed, it was trivial to create a new update package using old versions of Nordic's \texttt{nrfutil} program, activate DFU mode and then transfer the package using Nordic's Android DFU app.

\renewcommand\thesubsubsection{\thesubsection.\arabic{subsubsection}}

\subsection{Non-intrusive attack}
\label{sec:combining_exploits}

By combining the above exploits, a new, off-the-shelf TLL05A smart padlock can be  taken over, without registration with the manufacturer, and be given an over-the-air firmware update to turn it into a \droplock. This was first discussed in \cite[\S~III.D]{KerrisonDroplocksiThings2022}, and is based on the exploits detailed in this section.

\subsection{Specific recommendations}
\label{sec:specific_recommendations}

The descriptions of defenses, vulnerabilities and exploits in this section show how a successful over-the-air \droplock attack was made possible. While in totality, the attack is far from trivial, some significant improvements to the defenses in such devices are possible. Specific recommendations would include:

\begin{itemize}
  \item Avoid AES-ECB encryption and instead use a more robust AES such as AES-GCM or an alternative such as ChaCha.
  \item Use Diffie-Hellman to agree a key to encrypt HTTP API payloads, turning API snooping into a dynamic, per-session attack rather than a static global one.
  \item Do not rely on obscurity of the key derivation function to ensure reliance on cloud when establishing session keys.
  \item Implement mutual authentication between lock and cloud prior to establishing a session key to prevent offline (cloud-less) sessions, or support two modes of operation, with some activities requiring cloud, others not.
  \item Keep the processing of fingerprint images within an isolated environment so that they cannot be accessed in their original form (discussed further in the following section).
\end{itemize}

These recommendations may not apply directly to other smart locks, nor are they in themselves novel, but should be taken into consideration during the design and implementation process.

\subsection{On smartphone biometric security}
\label{sec:assessment_android}

To conclude this section, we refer to the Android 12 Compatibility Definitions Document (CDD) treatment of biometric sensors~\cite{AndroidBiometricSpecs} and assess whether the target lock was better or worse protected than what is required of many modern smartphones. Three classes of biometrics are referenced by the specification:  class 1, class 2 and class 3, with a higher class number implying stronger security. Much of the strength relates to biometric performance. For example minimizing the acceptance of spoof, impostor or false biometric samples, such as those that might be collected by a \droplock and then reproduced~\cite{Levalle2020}. However, the security of the biometric pipeline is also consideration, wherein raw biometric data cannot be extracted from the pipeline, or false data injected directly into it.

The CDD stipulates requirements for sensors depending on which class they are to be treated as, along with that functionality and behavior can be provided using the authorization of a sensor of such a class. Of particular interest to this work are remarks regarding chain of trust and use of a Trusted Execution Environment (TEE). At class 1, the CDD requires that new biometric data can only be added once trust has been established through other authentication factors. While we interpret this as an assurance that the biometric data represents the correct user, we note that it may also make it less likely to collect others' data inadvertently.

In class 2 devices, the CDD requires biometric matching to performed in a TEE or equivalent, preventing the raw biometric data from being leaked or tampered with. Complementary to this is the requirement that biometric data must not be accessible to the application processor outside the TEE. In the case of the target lock, the fingerprint reader chip does not operate with any isolation in force, and will release the biometric data unencrypted over UART. Were this not the case, the attacks presented in this section would be moot. Class 3 goes further still by requiring hardware-based protection of encryption keys. This is significantly beyond the capabilities observed on the target lock.

To summarize, the fingerprint sensing capabilities of the target lock closest resembles a class 1 sensor, intended for convenience rather than security. At this level, not only is the security of the authentication system relatively weak, so too is the protection of the biometric data, and we have shown why this poses a risk beyond the intended application.

\section{Applicability to other devices}
\label{sec:otherdevices}

Thus far a single smart padlock model has been used as a \droplock demonstrator, and while that is a sufficient risk to motivate putting mitigations in place, we must also seek to understand the scale of the problem. We do this in two ways: Assessing several other fingerprint-reading smart padlocks to estimate their vulnerability to a similar attack, and discussing other fingerprint-reading devices to identify ways in which they may also play a part in a biometric theft attack. Candidate devices would all need to possess some form of wireless communication capability, such as Wi-Fi or Bluetooth.

\subsection{Smart padlock survey}
\label{sec:smartlock_survey}

We acquired a selection of other lock models to perform a broader assessment. Locks were selected based on their availability at the time of purchase, seeking to avoid models that appeared to be re-badged versions of others already selected (based on product photographs). While the obtained devices have different shapes, weights, strengths, connectivity and application support, they are all padlocks that possess fingerprint readers and wireless connectivity. Ergo, if the firmware on the device can be reprogrammed, then there is potential to extract fingerprint data from the reader and transmit it over whatever radio is present in the device.

Hacking the original \droplock was a time-consuming process, but
once the DFU update method was cracked, the attack became significantly easier to perform. We do not seek to recreate this process in full for the surveyed locks, but rather perform an initial investigation into the likelihood that these devices could be similarly compromised.

For each lock, we assess the following aspects of its design and implementation, suggesting a rating of its security level in each of these areas. As with any such assessment, future discoveries may significantly change these, so this is only a point-in-time assessment constrained by time, resources and current knowledge.

The areas assessed are:
\begin{itemize}
  \item \textbf{Disassembly}: A more secure lock is harder to disassemble, ideally being impossible without causing visible irreparable damage.
  \item \textbf{Interface}: Easily accessible JTAG, SWD or SPI flash pads may make dumping, analyzing or modifying device firmware simpler to carry out. The most secure device would not include these post-production, or they would be very difficult to attach to.
  \item \textbf{Debug}: Regardless of interface accessibility, the most secure devices will have debug or external access to flash memory disabled, with no known workarounds.
  \item \textbf{DFU}: If a DFU package can be obtained and determined to be protected with PKI signatures, then replacing the firmware, particularly using FOTA, may be significantly harder for such a device.
\end{itemize}
We rate each of these areas based on the difficulty in leveraging them to perform a \droplock style attack, with low (L) being trivial, medium (M) requiring some skill or special equipment, and high (H) being highly complex or so-far unachievable. We exclude an assessment of whether each fingerprint chip prevents uploading whole images, as this would was not feasible to conduct with available equipment and resources at the time.

\begin{table*}
  \begin{center}
  \caption{Summary of security assessment of surveyed devices}
  \label{tab:locklist}
  \small
  \begin{tabular}{| c | c | c | c | c | l |}
  \hline
    \textbf{ID} & \textbf{Brand} & \textbf{Model} & \textbf{Source} & \textbf{Price} & \textbf{Notes} \\
  \hline
  \textbf{TLL05A} & Tapplock & Lite & Retail & \$50 & The \droplock demonstrator. \\
  \hline
  \textbf{TL203A} & Tapplock & One+ & Retail & \$100 & \\
  \hline
  \textbf{TY} & --- & --- & AliExpress & \$42 & Tuya compatible. \\
  \hline
  \textbf{MJ} & --- & --- & AliExpress & \$22 & Xiaomi Mijia compatible. \\
  \hline
  \textbf{TT} & --- & --- & AliExpress & \$30 & TTLock platform-based. \\
  \hline
  \end{tabular}
  \end{center}
\end{table*}

\begin{figure}
  \centering
  \includegraphics[width=.95\linewidth,trim={2cm 3cm 2cm 5cm},clip]{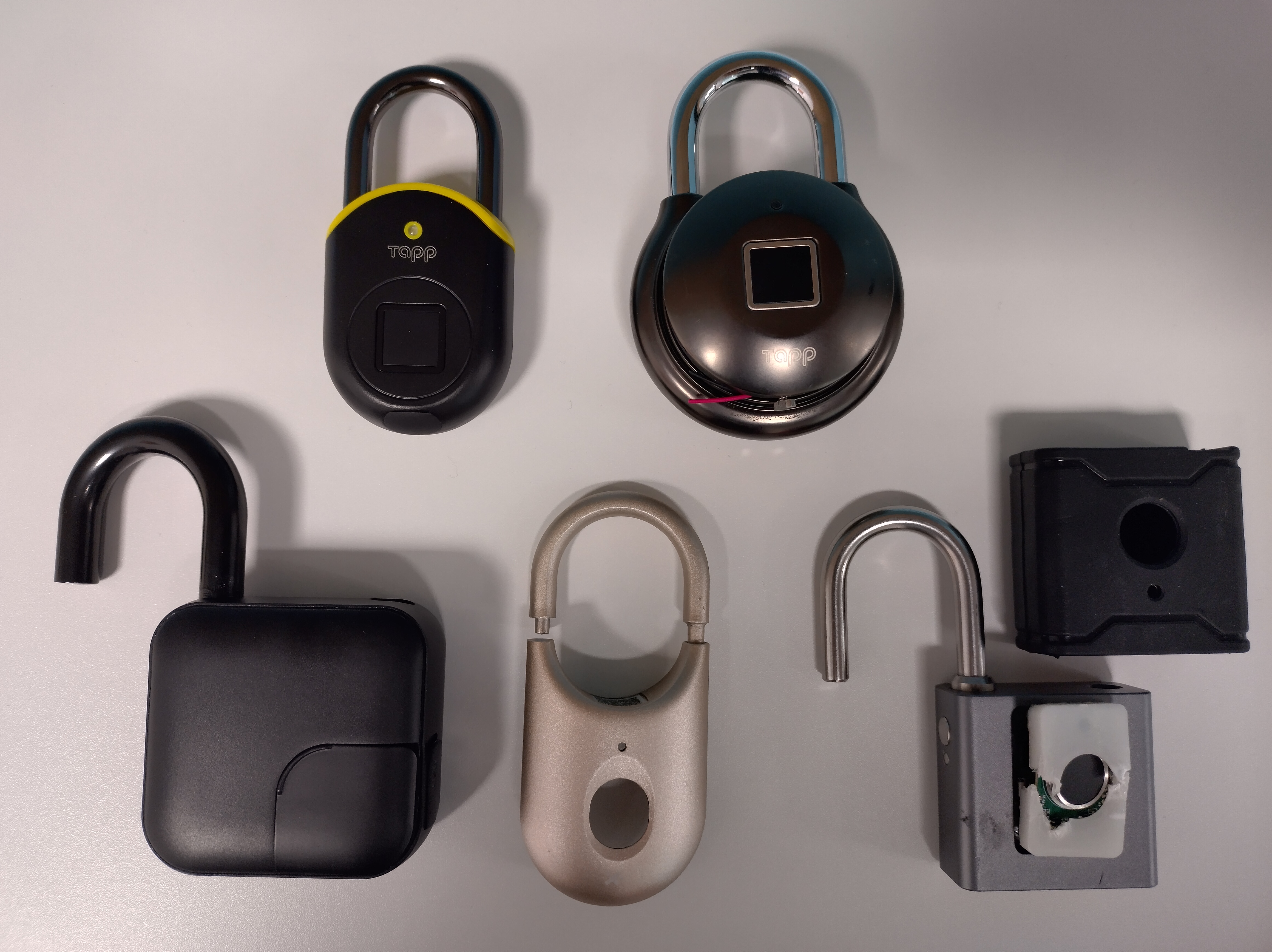}
  \caption{From top-left, clockwise: TLL05A, TL203A, TT, MJ and TY locks.}
  \label{fig:lock-set}
\end{figure}

The locks surveyed are detailed in \cref{tab:locklist} and pictured in various states of disassembly in \cref{fig:lock-set}. Those purchased from AliExpress have no, or confusing branding, typically featuring branding relating to the platform they are compatible with, though the lock itself is effectively no-name. For identifying the locks, we have used the branded product code, or a self-determined identifier based on some unique attribute. All of these locks use Bluetooth and have a fingerprint reader.

Tapplock devices are supported by Tapplock's own app and cloud. Tuya\footnote{Tuya: \url{https://www.tuya.com/}} is a Chinese smart device platform that provides libraries that OEMs can integrate into their products for compatibility with the platform and associated apps.  Mijia and TTLock are similar in nature, with Mijia being a platform from Xiaomi (known for smartphones and other smart devices) and TTLock\footnote{TTLock: \url{https://www.ttlock.com/}} focusing purely on the smart lock ecosystem.

\begin{table*}
  \begin{center}
  \caption{Summary of security analysis of surveyed devices}
  \label{tab:locksurvey}
  \small
  \begin{tabular}{| c | c | c | c | c | L{0.41\linewidth} |}
  \hline
    \textbf{Device} & \textbf{Disassembly} & \textbf{Interface} & \textbf{Debug} & \textbf{DFU} & \textbf{Notes} \\
  \hline
    \textbf{TLL05A} & L & M & L & L & Easy to disassemble/reassemble. Debug pads are small but labelled and unprotected. Only CRC16 on firmware images. \\
  \hline
    \textbf{TL203A} & H & L & L & L & Very difficult to disassemble non-destructively. Once inside, interfacing is comparatively easy. Only CRC16 on firmware images.\\
  \hline
    \textbf{TY} & M & L & H & ? & Appears to have flash/JTAG interface but could not detect. \\
  \hline
    \textbf{MJ} & M & L & H & ? & Undetectable over SWD debug pins. \\
  \hline
    \textbf{TT} & L & H & H & ? & RF-cover is glued, but rubber covering will mask any damage. MCU pins are epoxy-coated. \\
  \hline
    \textbf{\emph{Notes}} & \multicolumn{5}{L{0.85\linewidth}|}{\textit{Firmware packages were not obtained for some devices, therefore their DFU security was not assessable.}} \\
  \hline
  \end{tabular}
  \end{center}
\end{table*}

In the initial \droplock attack work, soldering and an ST-Link debugger were used to gain access to the device. In this survey, we introduced new tools to both speed up the work and reduce the risk of damage to devices. We used a series of pogo pin assemblies, combined with electrical tape to mask off areas of circuits if needed, to help attach to debug pads or pins solder-free. Additionally, a Raspberry Pi 4B was used alongside the ST-Link to provide an alternative debug interface when attempting to communicate with the various types of debug ports that were identified.

\Cref{tab:locksurvey} summarises our findings after assessing the devices. We were not as successful at accessing the firmware of other locks as we had been with the TLL05A. Some amount of luck may be to thank for our initial discoveries being based around this device. However, based on the extent of information gathering done, the locks do have varying degrees of protection in different areas.

\begin{figure}
  \centering
  \includegraphics[width=.95\linewidth,trim={2cm 4cm 2cm 7cm},clip]{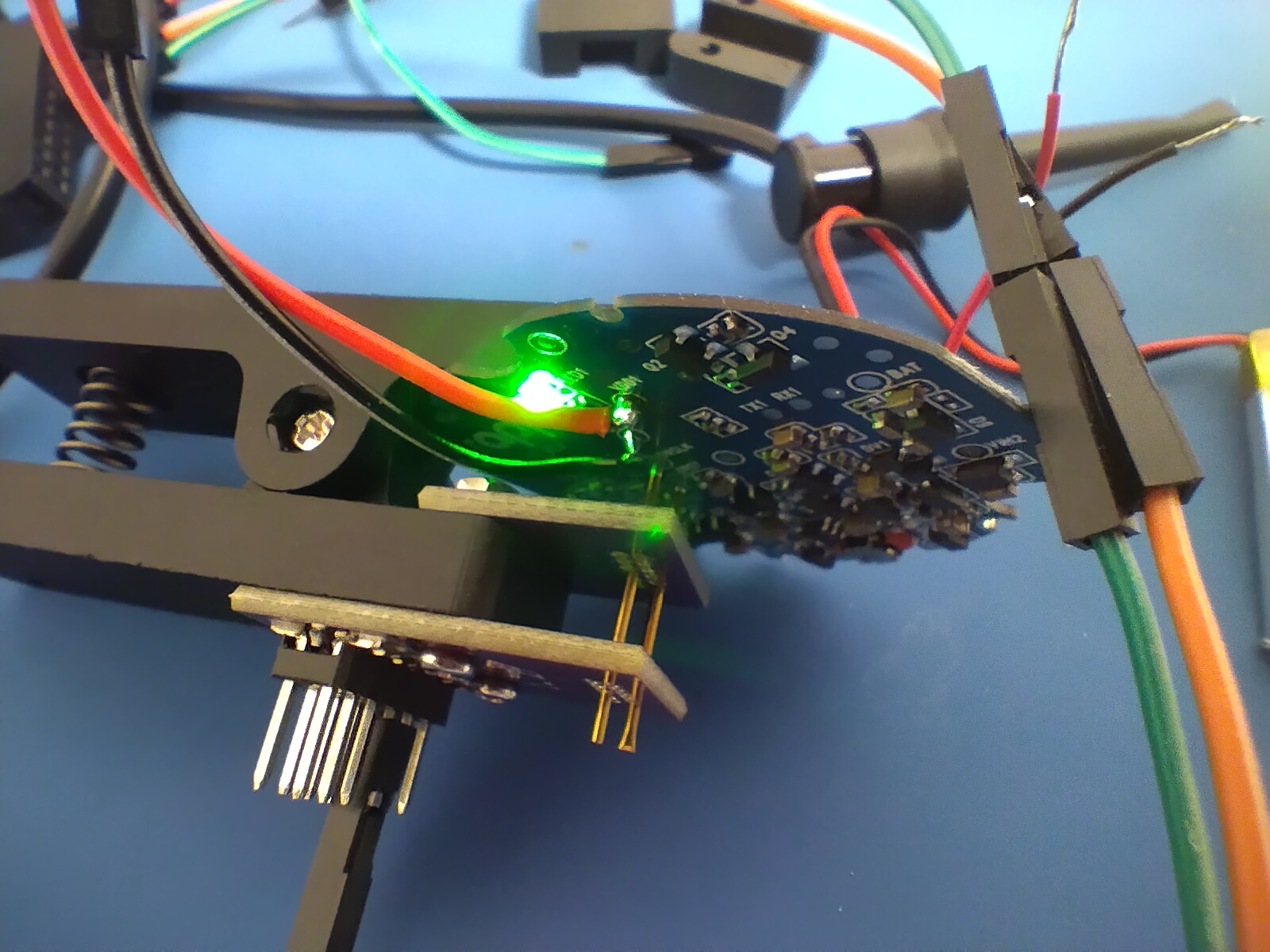}
  \caption{TL203A PCB with debug attachment via spring-loaded pogo pins.}
  \label{fig:tl203apogo}
\end{figure}

The TL203A is significantly harder to disassemble than the cheaper TLL05A from the same manufacturer, requiring drilling to remove the back of the lock. However, once inside, the same lack of debug protection and relative ease of interfacing make it possible to dump firmware and debug the device. \Cref{fig:tl203apogo} shows the TL203A rigged for debug, using a combination of soldered power/ground reference cables and a pogo pin clip on the device's debug clock and data pads.

We observe from a brief analysis of the TL203A's firmware and API data that its encryption implementation is different to the TLL05A, but believe that it could also be reverse engineered in a similar way, given sufficient time. This is supported largely by the fact that the BLE chips are the same and the DFU packages both lack any signature-based protection. We also note that the fingerprint chip on this device is labelled ID808+, which bears a striking similarity to the DF Robot device~\cite{ID809Source} used as reference for the fingerprint reader chip command protocol that was reverse engineered in \cite[\S III.B]{KerrisonDroplocksiThings2022}.

The TT lock appeared to have good physical construction, but the PCB is accessible behind a glued-in plastic cover. A protective rubberized cover could help mask any physical damage to the lock from disassembly and reassembly. The ML and TY locks are also relatively easy to get inside, but some evidence of this (scratches or dents) may remain afterwards. A prepared attacker could unlock any of the unbranded locks once their panels were removed, although this is not a concern in a \droplock attack. We note that unlike the Tapplock devices, which need to be woken by pressing some kind of button first, the TY, MJ and TT locks all wake when skin makes contact with their sensor. It is slightly easier to perform an attack on a victim if this wake-method can be used.

A variety of application MCUs were observed, including Nordic NRF51822, Telink TLSR8251, Beken BK3431 and Realtek/RealMCU RTL8762. Fingerprint chips include GigaDevices GDFFPR variants, Edward/IDWorld ID808+ and BYD BF5622A. This variation frustrates an attacker by requiring a different debug interface, SDK and system knowledge each time, assuming debug and/or reprogramming are even possible. More drastic approaches, such as de-soldering chips, may also aid in the hacking process, but we did not have the means to attempt this.

This survey of devices shows that varying level of protections are in place, mitigating different kinds of potential vulnerabilities. While defenses against disassembly may also be desirable for physical protection of the locking function, the other defenses mainly serve to protect the device from reverse engineering and reprogramming. Amongst the four additional locks, the TL203A was found likely to also be vulnerable to a \droplock conversion attack. Ultimately, any device will be convertible with sufficient effort, so the trade-off to be struck is the difficulty of doing so versus the likelihood of \droplock attack being successful, giving consideration to the value targeted individuals.

\subsection{Smart door locks}
\label{sec:doorlocks}

Unlike padlocks, door locks are fixed, either to the door, or to the frame of the door that they protect. We distinguish these from door entry systems used in enterprises, which have been in use for some time, whereas smart door locks have only entered the market in recent years.

For a \droplock-like attack to be feasible with these devices, the lock must be hacked in-situ, without triggering any alarm or anyone noticing. This is significantly more challenging than preparing a \droplock. Alternatively, if a target lock is identified, an attacker might purchase the same model, hack it, and assuming a FOTA process exists and is vulnerable to compromise, prepare an update that can be sent to the in-situ lock. Gaining adequate access to this unnoticed would still be a challenge.

While it does seem that such an attack would still be possible,
the likelihood of any one door lock being compromised in this way is less, and the lack of portability reduces attack flexibility. However, lock manufacturers should still be mindful of protecting fingerprint data in such devices, as devices that might connect to owners' smart home networks may in the future be found vulnerable to being turned en masse into fingerprint harvesters, which would cause significant reputational damage to the affected vendors.

Finally, we note that mounting an attack from a fixed device such as a door lock may help the attacker in selecting victims --- targeting those who use the particular entry-way --- rather than hoping a suitable target picks up a \droplock.

\subsection{Do it yourself}
\label{sec:diy}

We earlier posited that hacking an existing, recognized brand would help a \droplock attack be more successful. However, smart padlocks are thus far less mainstream than fixed smart door locks, which at the time of writing have products from well-known lock brands such as Yale and Assa Abloy. Additionally, USB thumb drives have many brands and styles, and yet people will recognize their function and be able to use any of them. As such, a custom-made \droplock may be feasible to produce if the time, skill and cost investment in making it is lower than hacking a COTS device.

We inquired for quotes from a number of smart lock OEMs, to determine if we could obtain a sample quantity of locks along with the ability to modify the firmware through an SDK. We also indicated an intention to purchase in bulk if the proof-of-concept was successful, in-line with any reasonable product development process.

The outcome of this inquiry was that none of the responding vendors of pre-assembled PCBs (PCBAs) were willing or able to offer firmware customization. This may be due the software and hardware supply chain, wherein we note the following:

\begin{itemize}
  \item Devices with a separate fingerprint chip may have a higher bill-of-materials and allow more feature customization, but the fingerprint handling firmware may not be under device manufacturer's control.
  \item Many devices integrate with smart device ecosystems (e.g. Tuya, TTLock, Mijia), wherein some libraries come from the platform provider, and device manufacturers may be unable to make an SDK for their product's firmware available due to 3rd party licensing restrictions.
\end{itemize}

While most OEMs could refuse to provide the firmware sources or SDK, that does not guarantee that \textit{all} OEMs would do so. As such, a determined attacker could expect to find a source for manufacturing their own \droplock{s} without having to create the whole product end-to-end themselves. And if not, existing designs could be used as reference for creating their own and developing a suitable firmware from scratch, with support from the chip-maker's SDK.

If this is the case, then even legitimate manufacturers can only do so much to protect users from a \droplock, as protecting genuine products may be insufficient. Additional steps may need to be taken to prove the genuine nature and purpose of a product to users, beyond simple visual cues, in real-time. This also requires increased awareness from users.

\section{Recommendations}
\label{sec:recommendations}

\begin{figure*}
  \centering
  \includegraphics[width=0.85\linewidth]{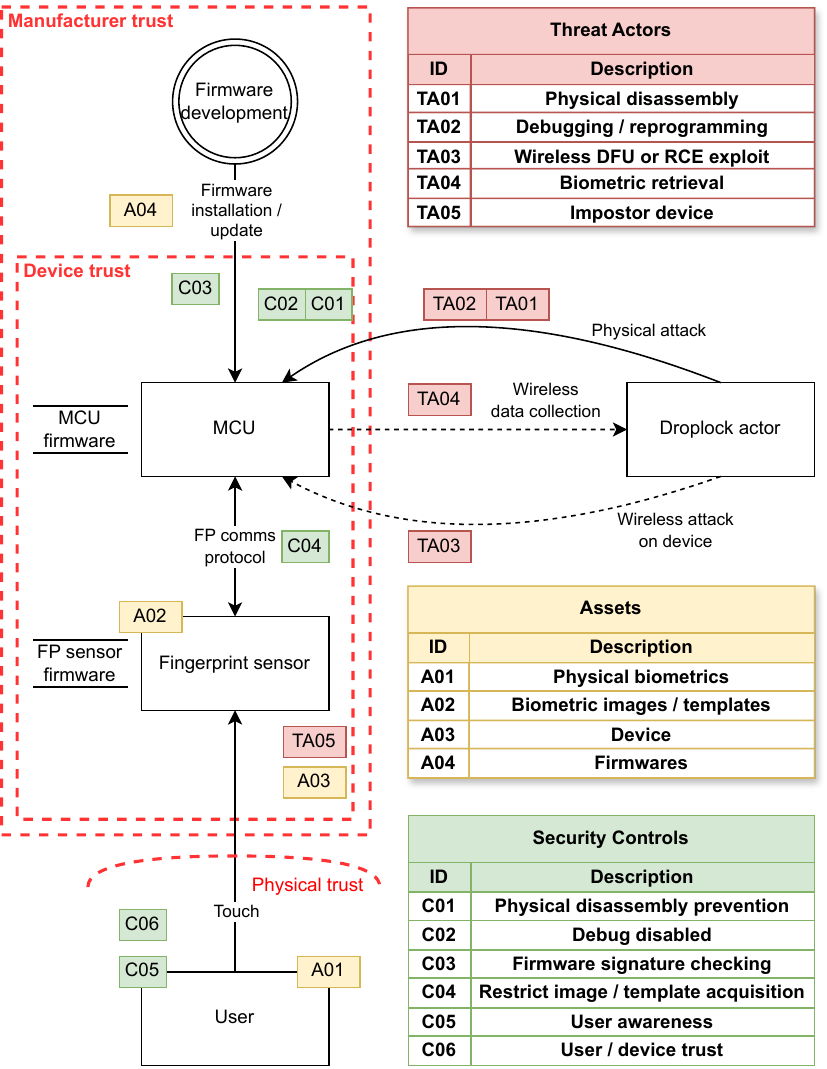}
  \caption{A data-flow / interaction-oriented threat model for threats, assets and controls relating to \droplock types of attacks.}
  \label{fig:threatmodel}
\end{figure*}

In \cref{fig:threatmodel} we present a threat model for a \droplock attack, labelling key threat actors, \texttt{TA}, assets, \texttt{A} and security controls \texttt{C} amongst data flows and interactions between manufacturer, IoT device, attacker and user. The goal of biometric data retrieval (\texttt{TA04}) can be achieved via physical attack (combining \texttt{TA01} and \texttt{TA02}), wireless attack (\texttt{TA03}) or use of an impostor device (\texttt{TA05}). The assets involved and our recommended security controls to mitigate the risks, are discussed herein.

\subsection{Make devices tamper-evident}

Physical protection (\texttt{C01}) of devices (\texttt{A03}) against tampering (\texttt{TA01}) is desirable to preserve their primary function, but it is also useful in preventing an attacker from gaining access to reprogram the device (if wireless DFU is otherwise secure). If physical disassembly is possible, it should at least be evident that this has been done, to give users an opportunity to raise suspicion about the device.

\subsection{Disable debug}

Disabling debug capabilities (\texttt{C02}), especially on production models, makes reverse engineering and reprogramming of devices significantly harder, delaying both development and deployment of a \droplock.

\subsection{Use PKI signed firmware updates}

\Cref{sec:remote} showed how inadequately protected DFU processes can be used to replace device firmware with malicious versions (\texttt{TA03}). Chip-makers provide PKI (firmware signing) solutions (\texttt{C03}) in their SDKs that should be used to protect firmware integrity (\texttt{A04}). Using old SDK versions may mean such features are unavailable.

\subsection{Prevent image upload}

Biometric data scanned by the fingerprint sensor (\texttt{A02}) should not be made available to other components in the smart device. Where this feature is configurablem, production devices should disable this capability (\texttt{C04}). It may be acceptable to extract template data, however, even these can be used to reproduce prints that may fool some sensors~\cite{Cappelli2007,Arakala2011}. We note that adhering to criteria similar to a class 2 biometric device, per Android's CDD, as discussed in \cref{sec:assessment_android}, would address this concern.

\subsection{Increase user awareness}

A community effort should be made to educate potential IoT device users (i.e. everyone) that an untrustworthy device could affect their security (\texttt{C05}). Aside from fingerprint theft, these devices could be bugged, used as wireless attack vectors or another invisible threat.

\subsection{Establishing user trust}

Unlike PCs or smartphones, there are few ways for a smart device to make known its integrity, principally due to the lack of a screen. It may be discernible upon trying to wirelessly connect with such a device, but by that time, it may be too late, as the victim may have picked up the device and touched the sensor. Building verification into a device's wireless capabilities would help to mitigate this (\texttt{C06}). It would also combat counterfeit devices, including custom-made \droplocks copying recognizable products (\texttt{TA06}). It would need to be accompanied by a security attitude of ``scan first, then interact'', and so must lean upon user awareness (\texttt{C05}) as well.

Mitigations in the former category alone are not sufficient protection, as a self-made \droplock (\cref{sec:diy}) would not be affected by them. This makes the user-focused mitigations equally important, although a trade-off should be made between the cost and effort of implementing all of these mitigations, and the likelihood of these kinds of attacks actually being carried out.

\subsection{Supply chain security integration}
\label{sec:supplychain}

Although not depicted in \cref{fig:threatmodel}, we wish to emphasize that device security is dependent on more than the manufacturer of the end-product. The supply chain is important too, including the robustness of the security features built into any chips used, the design of the PCBAs and the capabilities of the IoT cloud ecosystem with which the device communicates.

\section{Conclusion}
\label{sec:conclusion}

This paper has extended the understanding of the feasibility of performing \droplock style attacks from a technical standpoint. The attack is previously demonstrated as a proof-of-concept with simple hardware, then on a reverse-engineered COTS device, is ultimately achieved via remote firmware update of a brand new, unregistered / unenrolled COTS device. A detailed analysis of the vulnerabilities and exploit steps that can be used against the targeted lock is given, along with mitigation recommendations. A selection of other commercially available smart padlocks are then assessed with the same attack method in mind, with \(40\%\) of the surveyed devices likely to be vulnerable to \droplock conversion, albeit of the same brand, and the others possibly vulnerable if more sophisticated approaches were attempted.

Construction of a self-made \droplock was discussed, which may become a viable approach for a suitably motivated attacker if commercially available smart padlocks have enough countermeasures in place. General recommendations of such countermeasures were given, including the need for better user-awareness of the potential risks of interacting with biometrics-enabled IoT devices. Additional human behavioral research, standards for user-awareness and new methods of establishing device-user trust, were proposed as future work.

Having provided deeper technical insight into this scenario, a logical next step is to determine how susceptible users of various profiles would be to such an attack. This information would better motivate the effort required in responding to this threat. It may also be beneficial to consider susceptibility to non-biometric interactions, such as scanning QR codes found on dropped devices. Addressing unsolved recommendations proposed in \Cref{sec:recommendations} would also be beneficial to the cybersecurity community, along with a cost/benefit analysis of the solutions with respect to the likelihood and impact of such attacks.

\section*{Funding acknowledgment}

This research was funded by James Cook University internal research grant IRG20210022.

\bibliographystyle{unsrtnat}
{
  \small
  \bibliography{paper}

\begin{thebibliography}{14}
\providecommand{\natexlab}[1]{#1}
\providecommand{\url}[1]{\texttt{#1}}
\expandafter\ifx\csname urlstyle\endcsname\relax
  \providecommand{\doi}[1]{doi: #1}\else
  \providecommand{\doi}{doi: \begingroup \urlstyle{rm}\Url}\fi

\bibitem[Kerrison(2022{\natexlab{a}})]{KerrisonDroplocksiThings2022}
Steve Kerrison.
\newblock {IoT Droplocks: Wireless Fingerprint Theft Using Hacked Smart Locks}.
\newblock In \emph{2022 IEEE International Conferences on Internet of Things
  (iThings) and IEEE Green Computing \& Communications (GreenCom) and IEEE
  Cyber, Physical \& Social Computing (CPSCom) and IEEE Smart Data (SmartData)
  and IEEE Congress on Cybermatics (Cybermatics)}, pages 107--112,
  2022{\natexlab{a}}.
\newblock
  \doi{10.1109/iThings-GreenCom-CPSCom-SmartData-Cybermatics55523.2022.00054}.

\bibitem[Bhor and Karia(2017)]{BhorPinning}
Mahesh Bhor and Deepak Karia.
\newblock Certificate pinning for android applications.
\newblock In \emph{2017 International Conference on Inventive Systems and
  Control (ICISC)}, pages 1--4, 2017.
\newblock \doi{10.1109/ICISC.2017.8068748}.

\bibitem[{Android Developers}(2022)]{AndroidPinning}
{Android Developers}.
\newblock Security with network protocols: Restricting your app to specific
  certificates, 2022.
\newblock URL
  \url{https://developer.android.com/training/articles/security-ssl#Pinning}.

\bibitem[Kotipalli and Imran(2016)]{KotipalliSrinivasaRao2016HA}
Srinivasa~Rao Kotipalli and Mohammed~A Imran.
\newblock \emph{Hacking Android}.
\newblock Packt Publishing, Limited, Birmingham, 2016.
\newblock ISBN 9781785883149.

\bibitem[{max-r-b}(2018)]{libcipher-weak}
{max-r-b}.
\newblock Insecure key storage: secrets are very easy to retreive, 2018.
\newblock URL \url{https://github.com/linisme/Cipher.so/issues/30}.

\bibitem[Menezes et~al.(2001)Menezes, van Oorschot, and Vanstone]{HAC}
Alfred~J. Menezes, Paul~C. van Oorschot, and Scott~A. Vanstone.
\newblock \emph{Handbook of Applied Cryptography}.
\newblock CRC Press, 2001.
\newblock URL \url{https://cacr.uwaterloo.ca/hac/}.

\bibitem[Shah et~al.(2018)Shah, Shah, and Kansara]{ProGuard}
Yash Shah, Jimil Shah, and Krishna Kansara.
\newblock Code obfuscating a kotlin-based app with proguard.
\newblock In \emph{2018 Second International Conference on Advances in
  Electronics, Computers and Communications (ICAECC)}, pages 1--5, 2018.
\newblock \doi{10.1109/ICAECC.2018.8479507}.

\bibitem[{NSA}()]{Ghidra}
{NSA}.
\newblock Ghidra.
\newblock URL \url{https://ghidra-sre.org/}.

\bibitem[Kerrison(2022{\natexlab{b}})]{KerrisonDroplockData}
Steve Kerrison.
\newblock {IoT Droplock} source code and demonstration.
\newblock James Cook University, Jun 2022{\natexlab{b}}.

\bibitem[{Android Source}()]{AndroidBiometricSpecs}
{Android Source}.
\newblock Android 12 compatibility definition: Biometric sensors.
\newblock URL
  \url{https://source.android.com/compatibility/12/android-12-cdd#7310_biometric_sensors}.

\bibitem[Levalle(2020)]{Levalle2020}
Yamila Levalle.
\newblock Bypassing biometric systems with {3D} printer and enhanced grease
  attacks.
\newblock Technical report, Dreamlab Technologies, 2020.
\newblock URL
  \url{https://dreamlab.net/media/img/blog/2020-08-31-Attacking_Biometric_Systems/WP_Biometrics_v5.pdf}.

\bibitem[{DF Robot}()]{ID809Source}
{DF Robot}.
\newblock {DFRobot\_ID809}.
\newblock URL \url{https://github.com/DFRobot/DFRobot_ID809/}.

\bibitem[Cappelli et~al.(2007)Cappelli, Lumini, Maio, and
  Maltoni]{Cappelli2007}
Raffaele Cappelli, Alessandra Lumini, Davide Maio, and Dario Maltoni.
\newblock Fingerprint image reconstruction from standard templates.
\newblock \emph{IEEE Transactions on Pattern Analysis and Machine
  Intelligence}, 29, 2007.
\newblock ISSN 01628828.
\newblock \doi{10.1109/TPAMI.2007.1087}.

\bibitem[Arakala et~al.(2011)Arakala, Horadam, Jeffers, and
  Boztaş]{Arakala2011}
Arathi Arakala, K.~J. Horadam, Jason Jeffers, and Serdar Boztaş.
\newblock Protection of minutiae-based templates using biocryptographic
  constructs in the set difference metric.
\newblock \emph{Security and Communication Networks}, 4, 2011.
\newblock ISSN 19390122.
\newblock \doi{10.1002/sec.205}.

\end{thebibliography}
}

\vfill

\end{document}